\documentclass[useAMS,usenatbib,twocolumn]{mn2e}

\usepackage[dvipdfmx]{graphicx}
\usepackage{amsmath,amssymb,natbib,siunitx,color}

\usepackage{color}

\input epsf
\usepackage{graphicx}
\newcommand{\beq}{\begin{equation}}
\newcommand{\beqa}{\begin{eqnarray}}
		  \newcommand{\eeq}{\end{equation}}
\newcommand{\eeqa}{\end{eqnarray}}

\newcommand{\lsim}{\lesssim}
\newcommand{\gsim}{\gtrsim}
\newcommand{\psim}{\mbox{\raisebox{-1.0ex}{$~\stackrel{\textstyle \propto}
{\textstyle \sim}~$ }}}

\newcommand{\lmk}{\left(}
\newcommand{\rmk}{\right)}

\newcommand{\lkk}{\left[} 
\newcommand{\rkk}{\right]}
\newcommand{\lla}{\left\langle}
\newcommand{\p}{\partial}
\newcommand{\rra}{\right\rangle}
\newcommand{\so}{M_\odot}
\newcommand{\mch}{{\cal M}}

\title[]{Coupling of Dual Mass-Transferring White-Dwarf Binaries as a Variable Gravitational-Wave Emitter
}

\author[N. Seto]{Naoki Seto
\\
Department of Physics, Kyoto University, 
Kyoto 606-8502, Japan
}

\date{\today}

\begin{document}
\maketitle
\begin{abstract}

We study evolution of a hierarchical four-body (2+2) system composed by a pair of mass-transferring white dwarf binaries.  Applying a simplified model around the synchronous state of  two inner orbital periods, we newly find that the four body system could settle down to a limit cycle with a small period gap. 
The period gap  generates an amplitude  variation of emitted gravitational waves, as a beat effect. Depending on model parameters, the beat period could be  1-10\,yr and a large amplitude variation might be observed by space gravitational-wave detectors.

\end{abstract}



\section{introduction}

Synchronization phenomena have been widely observed in various research fields, including  physics, chemistry and biology (Pikovsky et al. 2003). So far, sound waves had been a quite efficient messenger  for identifying synchronized states. For example, in 1665, Huygens discovered a  synchronization capture of two coupled pendulum-clocks emitting ticking sounds. The emergence of synchronized clapping is another well-known example related to sound waves (N{\'e}da et al. 2000).

After the detection of GW150914 by advanced-LIGO, gravitational wave measurements have rapidly become a powerful tool for physics and astronomy (Abbott et al. 2016).  In general, binaries are considered to be promising sources of gravitational radiation in broad frequency regime.  In  the context of gravitational wave astronomy, Seto (2018) studied possibility of synchronization capture for hierarchical four-body (2+2) system composed by two inner binaries (see also Breiter \& Vokrouhlick{\'y} 2018; Tremaine  2020 for resonant interactions between inner binaries). He pointed out that mass-transferring white dwarf (WD) binaries (AMCVn stars)  could be intriguing systems for realizing a synchronized state in the LISA band. This is because of the self-regulated nature of their  mass transfers. More specifically, in contrast to a binary effectively formed by two point masses (e.g. binary black holes),  the time-dependent mass transfer rate can efficiently soften the response of inner angular velocity to externally added torque.  Seto (2018) also found that,  after the synchronization capture, the luminosity of gravitational radiation will decrease significantly, due to the phase cancellation of the two coupled wave sources. Furthermore, a parasitic relation between the two binaries will be a likely outcome, and one of the two binaries seizes angular momentum  from the other, with the assistance of the synchronization.

LISA  is expected to detect $10^3$-$10^4$ isolated  mass-transferring WD binaries  ({Nelemans}, {Yungelson} \& {Portegies Zwart} 2004). One might be further interested in the formation scenarios of  a hierarchical 2+2 system composed by two WD binaries.
In the main sequence stage, for  nearby solar-type (F and G) dwarfs,  the fraction of hierarchical 2+2 systems is estimated to be   $\sim 4\%$ of the total systems (Tokovinin 2014; see also Raghavan et al. 2010). The observed multiplicity fraction  is known to be  generally higher for more massive stars such as the O-type stars (Sana et al. 2013).  
However, during the stellar evolution,   we need significant shrinkages of the inner and outer orbits to make the compact 2+2 systems as studied in Seto (2018) and also in this paper. 

In the case of an isolated WD binary, the common-envelope (CE) phase is considered to be important for its orbital contraction.  But, its basic physical processes  are not well understood at present (for reviews see e.g.  {Iben \& Livio} {1993}; {Taam \& Sandquist} {2000).   Compared with binaries, the effects of the CE phases would be much more complicated for triple and quadrupole systems   (for triple systems, see e.g. de Vires et al. 2014, Toonen et al. 2016; Glanz \& Perets 2020). From a pessimistic perspective, there is a possibility that, in many cases,  the 2+2 structure  might not be maintained during the CE phases (as naively speculated from Glanz \& Perets 2020).  
But, considering the large uncertainties,   it would be currently  difficult to make  solid discussions on the roles of the CE phases for 2+2 systems.

In  this paper, as in Seto (2018), we rather concentrate on the evolution of coupled two mass-transferring WD binaries only around the synchronization states. 
To extract essential degrees of freedom and keep robustness of our discussion, we use a very simple model for conservative  mass transfers, based on Paczy{\'n}ski (1967) and Paczy{\'n}ski \& {Sienkiewicz} (1972).  We exclude all the details added in later studies (e.g. spin effects). 

 In this paper, as contrasted  to the synchronization capture studied in Seto (2018),   we newly report the existence of a limit cycle for gravitationally coupled two WD binaries.    In this state,  the two binaries keep slightly different angular velocities and  periodically change gravitational wave luminosity as a beat effect.  Depending on model parameters, the beat period could become 1-10\,yr and might be actually observed by space gravitational-wave interferometers such as LISA, TianQin and Taiji.

\section{basic equations}
\begin{figure}
 \includegraphics[width=7.5cm,angle=270]{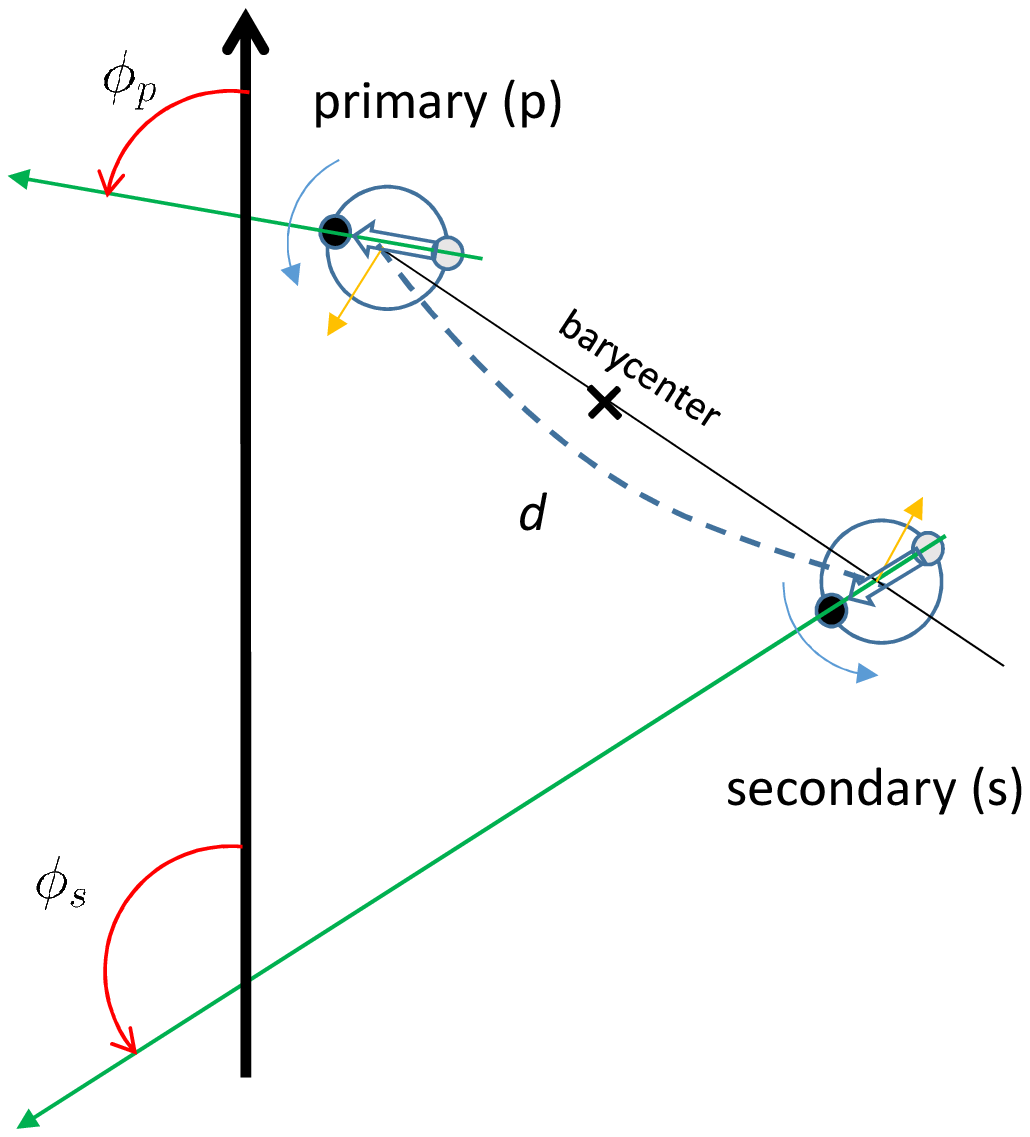}
 \caption{The geometry of dual mass-transferring WD-WD binaries (p: primary and s: secondary) aligned on the same plane. The gray circles show the lighter components in the two binaries ($m_{\rm p2}$ and $m_{\rm s2}$) corresponding to  the donors of  mass transfers.  The orientation  angles $\phi_{\rm p}$ and $\phi_{\rm s}$ are measured relative to the fixed direction (thick solid line).    We define the relative angle $\Delta\equiv \phi_{\rm s}-\phi_{\rm p}$. The outer orbital distance  is $d$ and the inner semi-major axes are given by $a_{\rm p}$ and $a_{\rm s}$. The three orbits are assumed to be circular. }
\end{figure}

We study evolution of a hierarchical four-body system composed by two mass-transferring WD binaries, as illustrated in Fig. 1. We assume that the three (one outer and two inner) orbits are circular and aligned on the same plane (see also Fang, Thompson, \& Hirata 2017; Hamers \& Lai {2017}; Fragione \& Kocsis {2019} for possible effects of inclination).  
We put the total masses of the two binaries by $M_{\rm p}$ (p: primary) and $M_{\rm s}$ (s: secondary) with $M_{\rm p}>M_{\rm s}$,  and denote their mutual distance by $d$ (see Fig. 1).  
We use the angles $(\phi_{\rm p},\phi_{\rm s})$ to represent the orientations of the binaries relative to a fixed direction (Fig. 1). 
Then, we define the relative orientation angle $\Delta\equiv \phi_{\rm s}-\phi_{\rm p}$ which plays an  important role in this paper.

This section is organized in the following order.  In \S 2.1, we describe our model  for the inner orbits.  In \S 2.2, we mention coupling between the two binaries and present resonant terms.  In \S 2.3, we summarize differential equations used for our numerical calculations. Then, in \S 2.4, we  derive some expressions that will be useful  to interpret our numerical results.

\subsection{INNER BINARIES}

In this subsection, we mainly discuss the primary binary, but we will apply  our results equally  to the secondary binary, after changing subscripts.

We put individual masses of the primary binary by $m_{\rm p1}$ and $m_{\rm p2}$ with $m_{\rm p1}>m_{\rm p2}$ and $M_{\rm p}=m_{\rm p1}+m_{\rm p2}$. The reduced mass is given by $\mu_{\rm p}\equiv m_{\rm p1}m_{\rm p2}/M_{\rm p}$  and the mass ratio by $q_{\rm p}\equiv m_{\rm p2}/m_{\rm p1}<1$.  
With its semi-major axis $a_{\rm p}$, the orbital angular velocity of the primary is given by 
\beq
n_{\rm p}\equiv \dot{\phi}_{\rm p}=\lmk{{GM_{\rm p}}/{a_{\rm p}^3}}\rmk^{1/2},
\eeq 
and its quadrupole moment is expressed as 
\beq
\mu_{\rm p} a_{\rm p}^2=G^{2/3}\mch_{\rm p}^{5/3}n_{\rm p}^{-4/3}
\eeq
with the chirp mass
$\mch_{\rm p}\equiv \mu_{\rm p}^{3/5} M_{\rm p}^{2/5}$. 

The orbital angular momentum of the primary binary is written by 
\beq
J_{\rm p}=\mu_{\rm p} (G M_{\rm p} a_{\rm p})^{1/2}=\mu_{\rm p} a_{\rm p}^2 n_{\rm p}. \label{jp}
\eeq
In the following, we deal with conservative mass transfer (i.e. $\dot{M}_{\rm p}=\dot{m}_{\rm p1}+\dot{m}_{\rm p2}=0$). We will shortly explain a concrete model for the rate $\dot{m}_{\rm p2}$.  From Eq. (\ref{jp}), we have
\beq
\frac{{\dot J}_{\rm p}}{J_{\rm p}}=\frac12 \frac{{\dot a}_{\rm p}}{a_{\rm p}}+\frac{{\dot m}_{\rm p2}}{m_{\rm p2}} (1-q_{\rm p}).
\eeq

The binary emits gravitational radiation mainly at the wavelength $\lambda_{\rm p}=\pi c/n_{\rm p}$ (frequency: $n_{\rm p}/\pi$). If the coupling between the two binaries are negligible,  the rate of angular momentum loss is given by (e.g. {Maggiore} 2008)
\beq
({\dot J}_{\rm p})_{\rm gw}=Y_{\rm pp}=-\frac{32 Ga_{\rm p}^4 n_{\rm p}^5 \mu_{\rm p}^2}{5c^5}.\label{jgwp}
\eeq
The associated timescale is given by  
\beqa
{t_{\rm gw,p}}&\equiv& -\lkk \frac{({\dot J}_{\rm p})_{\rm gw}}{J_{\rm p}} \rkk^{-1}=\frac{5c^5}{32 G^{5/3} \mch_{\rm p}^{5/3} n_{\rm p}^{8/3}} \label{tgw}\\
&=&1.5\times 10^8 \lmk  \frac{\mch_{\rm p}}{0.133M_\odot}\rmk^{-5/3}\lmk\frac{n_{\rm p}}{\rm 0.005\,s^{-1}}\rmk^{-8/3}{\rm \,yr}.\nonumber
\eeqa

Next, we move to  discuss the mass transfer rate within each binary.  We follow Paczy{\'n}ski  (1967) and Paczy{\'n}ski \& {Sienkiewicz} (1972)  for the Roche lobe overflow within a white dwarf binary.
 For the equation of state of WDs, we use the polytropic model with the index 3/2 for non-relativistic degenerate gas,  and the mass-radius relation is given by ($\rm i=1,2$)
\beq
R_{\rm pi}=0.0126 R_\odot \lmk \frac{m_{\rm pi}}{1M_\odot}\rmk^{-1/3}\label{mr}
\eeq
(see e.g.,  Zapolsky \& Salpeter 1969 for a more detailed modeling).
From Eq. (\ref{mr}), we have $R_{\rm p2}>R_{\rm p1}$ and the lighter component of the binary is the donor with $\dot{m}_{\rm p2}<0$.  We estimate  its  Roche lobe radius by (Paczy{\'n}ski  1967)
\beq
R_{\rm Lp2}=\frac{2a}{3^{4/3}} \lmk \frac{m_{\rm p2}}{M_{\rm p}}   \rmk^{1/3}.\label{rl}
\eeq
The mass transfer is stable for the condition $d(R_{\rm Lp2}/R_{\rm p2})/dm_{\rm p2}<0$ (for $J_{\rm p}=$const) and  this can be simplified as $q_{\rm p}=m_{\rm p2}/m_{\rm p1}<2/3$ (see e.g. Paczy{\'n}ski  1967; Solheim {2010}).  We use the mass transfer  rate 
\beq
\frac{{\dot m}_{\rm p2}}{m_{\rm p2}}=-2n \lmk \frac{R_{\rm p2}-R_{\rm Lp2}}{R_{\rm p2}}  \rmk^3\theta(R_{\rm p2}-R_{\rm Lp2}) \label{dm}
\eeq
given by the competition between $R_{\rm Lp2}$ and $R_{\rm p2}$ with the step function $\theta(\cdot)$ (Paczy{\'n}ski \& {Sienkiewicz} 1972; Webbink 1984, see also Marsh, Nelemans \& Steeghs 2004). In fact, the step function plays no role for most of our numerical calculations below (except for \S3.2).  It should be also  noticed that, in our study, as long as our modeling is valid,   the compact accreter is not necessarily a white dwarf. {But, at least for isolated binaries,  LISA is likely to detect double white dwarf binaries much more than binaries including  neutron stars or black holes (see e.g. Nelemans et al. 2001).   }

The stable and self-regulated mass transfer (\ref{dm}) is crucially important for our study.  It  softens the response of angular velocity $n_{\rm p}$, against externally added torque, resulting in dynamically interesting phenomena.  
 During the mass transfer phase, the binary satisfies the relation
\beq
0<R_{\rm p2}-R_{\rm Lp2}\ll R_{\rm p2}.
\eeq 
Then, from Eqs. (1)(\ref{mr}) and (\ref{rl}) the donor mass $m_{\rm p2}$ is approximately given by the angular velocity $n_{\rm p}$ as
\beq
m_{\rm p2}\simeq 0.036 \lmk \frac{n_{\rm p}}{0.005\,{\rm s^{-1}}} \rmk M_\odot . \label{dom}
\eeq

For an isolated binary, at quasi-steady state of mass transfer $\ddot{m}_{\rm p2}\simeq0$, we have ({Gokhale}, {Peng} \& {Frank} 2007)
\beq
\frac{{\dot a}_{\rm p}}{a_{\rm p}}=-\frac23\frac{{\dot n}_{\rm p}}{n_{\rm p}}=-\frac23\frac{{\dot m}_{\rm p2}}{m_{\rm p2}}=\lmk1-\frac32q_{\rm p}  \rmk^{-1} t_{\rm gw,p}^{-1}. \label{iso}
\eeq
 In \S 3, we use this relation to set up the initial conditions for our numerical calculations.

\subsection{COUPLING BETWEEN TWO BINARIES}

We now discuss gravitational coupling between two binaries around the synchronization state $\dot{\Delta}=n_{\rm s}-n_{\rm p}\simeq0$. We extract the relevant resonant terms caused by the Newtonian tidal interaction and the gravitational radiation reaction. The former is the leading order term of the conservative effects and the latter is that of the dissipative effects.  Throughout this paper, we assume that the four-body system is in the near zone ($d\ll \lambda_{\rm p}\simeq\lambda_{\rm s}$), and  ignore the time retardation for the couplings. We also put aside short-duration terms that depend on rapidly changing angular variables.  But these terms might play certain roles in some cases.

Due to the Newtonian tidal interaction with the secondary,  the primary receives the following  resonant torque  
\beqa
T_{\rm p}&=&\frac{9G a_{\rm p}^2 a_{\rm s}^2 \mu_{\rm p} \mu_{\rm s}}{16d^5} \sin (2\Delta). \label{new}
\eeqa
This expression is consistent with Tremaine (2020).  The secondary receives the counter torque $T_{\rm s}=-T_{\rm p}$.  Given the conjugate structure of the variables, these terms are not directly related to the time variation of eccentricities ({Murray} \& {Dermott} 1999). 

Next, we deal with the coupling between the  two binaries due to the gravitational radiation reaction. In most situations, such effect is totally negligible. But, for our systems with $d\ll \lambda_{\rm p}\simeq\lambda_{\rm s}$, the coherent nature could be exceptionally workable. From the Burke-Thorne potential (Thorne 1969; {Burke} 1971; Maggiore 2008),  the radiational torque on the primary due to the secondary is estimated to be   
\beq
Y_{\rm ps}=-\frac{32 Ga_{\rm p}^2 a_{\rm s}^2 n_{\rm s}^5 \mu_{\rm p} \mu_{\rm s}}{5c^5} \cos(2\Delta). \label{yss}
\eeq
Similarly, the secondary receives the following torque 
\beq
Y_{\rm sp}=-\frac{32 Ga_{\rm s}^2 a_{\rm p}^2 n_{\rm p}^5 \mu_{\rm p} \mu_{\rm s}}{5c^5} \cos(2\Delta). \label{ysp}
\eeq

\subsection{EQUATIONS FOR NUMERICAL STUDIES}
In this subsection,  for a preparation of numerical calculations,  we summarize expressions provided so far. 
Hereafter, for notational conciseness, we put $n_{\rm p}=n_{\rm s}=n$ ($\lambda_{\rm p}=\lambda_{\rm s}=\lambda$), unless the difference between $n_{\rm p}$ and $n_{\rm s}$ should be clarified.

First, we write down the total torque for each binary.  From Eqs.  (\ref{jgwp})(\ref{new})(\ref{yss}) and (\ref{ysp}), we have 
\beqa
\frac{{\dot J}_{\rm p}}{J_{\rm p}}&=&\frac{Y_{\rm pp}+Y_{\rm ps}+T_{\rm p}}{J_{\rm p}}\\
&=&-\frac{1}{t_{\rm gw,s}} [F+\cos(2\Delta)-D\sin(2\Delta)]\label{djp}\\
\frac{{\dot J}_{\rm s}}{J_{\rm s}}&=&\frac{Y_{\rm ss}+Y_{\rm sp}+T_{\rm s}}{J_{\rm s}}\\
&=&-\frac1{t_{\rm gw,s}} [1+F\cos(2\Delta)+DF\sin(2\Delta)]\label{djs}
\eeqa
for the primary and secondary. 
Here we introduced the following two parameters that will become important in the rest of this paper
\beqa
F&\equiv& \lmk  \frac{\mch_{\rm p}}{\mch_{\rm s}}\rmk ^{5/3},\label{ff}\\
D&\equiv& \frac{45 c^5 n^{-5}}{2^9d^5}=\frac{45 }{512\pi^5} \lmk\frac{\lambda}{d}\rmk^5\\
&=&36.2 \lmk \frac{n}{\rm 0.005 \,s^{-1}} \rmk^{-5} \lmk \frac{d}{\rm 0.12\,AU}  \rmk^{-5}.\label{defd}\nonumber
\eeqa
In the absence of the coupling terms, we have  $F=\dot{n}_{\rm p}/\dot{n}_{\rm s}\ge 1$ for the  two angular speeds. To reduce the encounter speed $\propto (F-1)$ and thereby  enhance  dynamical interaction around the synchronization condition $\dot{\Delta}\sim 0$, we numerically study the cases with $0<F-1\ll 1$.   Meanwhile, the parameter $D$ represents the strength of the Newtonian torque relative to the radiative ones. Its prefactor $45/(512\pi^6)\sim3\times 10^{-4}$ is much smaller than unity.  Considering the requirement  $\lambda\gg d$, we mainly study the range $D\gsim 10$.  

We also need to take into account the orbital stability for the four-body system. We apply the stability criterion in Mardling \& Aarseth (2001) by considering an effective triple system composed the secondary binary (masses $m_{\rm s1}$ and   $m_{\rm s2}$) and the third body of the primary\rq{}s total mass $M_{\rm p}$.  Then, for $M_{\rm p}\sim m_{\rm s1}+m_{\rm s2}$,   we obtain the upper limit for the coupling parameter
\beq
D_{\rm max}\sim 6.0\times 10^8 \lmk \frac{n}{0.005 {\rm\, s^{-1}}}  \rmk^{-5/3}. 
\eeq 
In what follows, we examine the regime $D\ll D_{\rm max}$.

In our numerical calculations,  we trace the time evolution of the five  variables  $\Delta$, $a_{\rm p}$, $a_{\rm s}$, $m_{\rm p2}$ and $m_{\rm s2}$, using the five differential equations below.  From the balance of angular momenta, we have
\beq
\frac12 \frac{{\dot a}_{p}}{a_{\rm p}}+\frac{{\dot m}_{\rm p2}}{m_{\rm p2}} (1-q_{\rm p})=-\frac{1}{t_{\rm gw,s}} [F+\cos(2\Delta)-D\sin(2\Delta)],\label{djp2}
\eeq
\beq
\frac12 \frac{{\dot a}_{s}}{a_{\rm s}}+\frac{{\dot m}_{\rm s2}}{m_{\rm s2}} (1-q_{\rm s})=-\frac{1}{t_{\rm gw,s}} [1+F\cos(2\Delta)+DF\sin(2\Delta)].\label{djp3}
\eeq
From the definition of the relative angle $\Delta$, we have
\beq
{\dot \Delta}=\dot{\phi_{\rm s}}-\dot{\phi_{\rm s}}=\lmk \frac{GM_{\rm s}}{a_{\rm s}^3}  \rmk^{1/2}-\lmk \frac{GM_{\rm p}}{a_{\rm p}^3}  \rmk^{1/2}.\label{dd}
\eeq
In addition, we use Eq. (\ref{dm}) for the mass transfer rate $\dot{m}_{\rm p2}$ and a similar one for  $\dot{m}_{\rm s2}$.

\subsection{ENERGY EQUATION}
As in Seto (2018), an energy equation for $\Delta$ is useful to understand evolution of the coupled four-body system (see also Goldreich \& {Peale} 1968; {Murray} \& {Dermott} 1999 for another example).  Here, we briefly discuss the basic aspects of the energy equation. From Eq. (1), we have
\beq
{\ddot \Delta}={\dot n}_{\rm s}-\dot{n}_{\rm p}=-\frac32n \lmk \frac{\dot a_{\rm s}}{ a_{\rm s}}- \frac{\dot a_{\rm p}}{ a_{\rm p}} \rmk. \label{b1}
\eeq
In the right-hand side of this expression,  we dropped a correction of $O[(n_{\rm p}-n_{\rm s})/n]$.
  Then, using Eqs. (\ref{djp2}) and (\ref{djp3}),  we obtain
\beqa
{\ddot {\Delta}}&-&\frac{3n}{t_{\rm gw,s}} \lkk (1-F) (1-\cos2\Delta)+D(F+1)\sin2\Delta\rkk\nonumber \\
&=&3n \lkk \frac{\dot{m}_{\rm s2}}{m_{\rm s2}}(1-q_{\rm s})- \frac{\dot{m}_{\rm p2}}{m_{\rm p2}}(1-q_{\rm p}) \rkk. 
\eeqa
Multiplying $\dot \Delta$ and integrating with time, we obtain 
\beq
\frac12 {\dot \Delta}^2+ V(\Delta)=E(t),\label{et}
\eeq
where the potential $V(\Delta)$ is given by 
\beq
V(\Delta)\equiv  \frac{3n}{2t_{\rm gw,s}} \big[ (F-1) (2\Delta-\sin2\Delta)
 +D(F+1)\cos2\Delta\big] .\label{pot}
\eeq
Here, we ignored the time variations of the parameters $(n,F,D,t_{\rm gw,s})$, since we  are interested in a time period much shorter than $t_{\rm gw,s}$.
Similarly, the total energy $E(t)$ can be evaluated by
\beqa
 E(t)&=&  
 E(0) \nonumber \\
& & +3n\int_{0}^t dt \lkk \frac{\dot{m}_{\rm s2}}{m_{\rm s2}}(1-q_{\rm s})- \frac{\dot{m}_{\rm p2}}{m_{\rm p2}}(1-q_{\rm p}) \rkk {\dot \Delta}\label{et2}
\eeqa
with an integral constant $ E(0)$. This expression shows that the total energy $E(t)$ is changed by the mass transfers.   For numerical evaluation of $E(t)$, we exclusively  apply the left-hand side of Eq. (\ref{et}), and use Eq. (\ref{et2}) only for analytical studies.

 As mentioned earlier, we mainly analyze coupled binaries with $0<F-1\ll 1$ and $D\gg1$.  For such parameters, in Eq. (\ref{pot}), the local profile of the potential $V(\Delta)$ is dominated by the term $\propto D(F+1)\cos2\Delta$ with a small gradient $\propto 2(F-1)\Delta$.

\section{numerical results}

We now numerically study  the time evolution of  two mass-transferring WD binaries, gravitationally coupled at the distance $d$.   In this section,  we assign  various coupling parameters $D$, but unless otherwise stated, other conditions are identical (except for the last paragraph in \S3.4). More specifically,  we put $(M_{\rm p},M_{\rm s})=(1.0M_\odot,0.9M_\odot)$ and use the common initial conditions (at $t=0$); $\Delta=0$,  $n_{\rm s}=5\times 10^{-3}{\rm s^{-1}}$ and  $n_{\rm p}=(1+2\times 10^{-5})n_{\rm s}$.   We finely adjust the initial  donor masses $(m_{\rm p2},m_{\rm s2})$ to individually satisfy the third equality in Eq. (\ref{iso}) that is  originally given for an isolated binary. Roughly speaking, for $n_{\rm p}\sim n_{\rm s}\sim5\times 10^{-3}{\rm \,s^{-1}}$, we have $m_{\rm p2}\sim m_{\rm s2}\sim 0.036M_\odot$ (accordingly $\mch_{\rm p}\sim0.133\so$, $\mch_{\rm s}\sim0.127\so$,  $t_{\rm gw,s}\sim 1.6\times 10^8$\,yr and $F\sim 1.046$). 

\subsection{OVERALL PHASE EVOLUTIONS}

To begin with, we discuss the overall evolution of the phase difference $\Delta$ for the four different coupling parameters $D=0.1,24,25$ and 50.  Here the parameter $D=0.1$ is not comfortably large, considering the near zone condition $d\ll \lambda$. We use this run just for a comparison.

In Fig. 2, we show our  numerical results.  In the early stage $t\lsim 20000$\,yr, the coupling between the two binaries is non-resonant and inefficient.  Therefore, in this stage,  we will be able to make an extrapolation
\beq
\ddot{\Delta} \simeq \ddot{\Delta}_{\rm iso} . \label{ext}
\eeq
Here  $ \ddot{\Delta}_{\rm iso}$ is given by  $({\dot n}_{\rm s}-{\dot n}_{\rm p})$ for two isolated binaries as in Eq. (\ref{iso}), and its time variation can be neglected for the timescale under discussion. Integrating Eq. (\ref{ext}) twice and using the initial condition $\Delta(0)=0$, we obtain  a parabolic equation as an  approximation to $\Delta(t)$
\beq
\Delta_{\rm pb} (t)= \frac12  \ddot{\Delta}_{\rm iso} \times (t-2t_{\rm c})t . \label{pb}
\eeq
Here we defined the expected catch-up time   for the two angular speeds 
\beq
t_{\rm c}= \frac{(n_{\rm p}-n_{\rm s})_0}{ \ddot{\Delta}_{\rm iso}}\sim 2.6\times 10^4 {\rm \,yr}
\eeq 
using the initial velocity difference $(n_{\rm p}-n_{\rm s})_0$.  We also  denote the corresponding phase by  
\beq
\Delta_{\rm c}\equiv \Delta_{\rm pb}(t_{\rm c})\sim- \frac12\ddot{\Delta}_{\rm iso} t_{\rm c}^2\sim -40000.
\eeq
In fact, we set the initial difference $(n_{\rm p}-n_{\rm s})_0/n_{s0}=2\times 10^{-5}$ to have a large rotation cycles $|\Delta_{\rm c}|/(2\pi)\sim 10^4$.  Thus, around the critical epoch  $t\sim t_{\rm c}$,  we will be able to  suppress transient effects caused by our potentially artificial initial settings.

As shown in Fig.  2,  depending weakly on $D$, the time profile $\Delta(t)$ at $0<t\lsim t_{\rm c}$ is approximately given by the analytical expression (\ref{pb}).
For $D=0.1$, the coupling between the binaries are weak,  even around $t\sim t_{\rm c}$, and they merely pass through the resonant point ${\dot \Delta}=0$,  following the expression (\ref{pb}) still at $t>t_{\rm c}$.
In contrast, for $D=24, 25$ and 50, we have quite different profiles $\Delta(t)$ at $t\gsim t_{\rm c}$.  For $D=24$, the binaries are captured into a synchronization state  $\Delta\sim 40286$ (as discussed in \S3.2). Meanwhile, for $D=25$ and 50, the systems asymptotically show constant drifts ${\dot \Delta}\sim \rm const$. 

Seto  (2018) closely examined the success and failure of the synchronization capture (respectively corresponding to $D=24$ and  0.1 in Fig. 2).  However,  the  existence of a drifting solution was not reported at all. Therefore, in this follow-on paper, we mainly discuss  the drifting solutions (e.g. $D=25$ and 50), paying special attention to the boundary between the synchronization capture (e.g. $D=24$).

\begin{figure}
 \includegraphics[width=.95\linewidth]{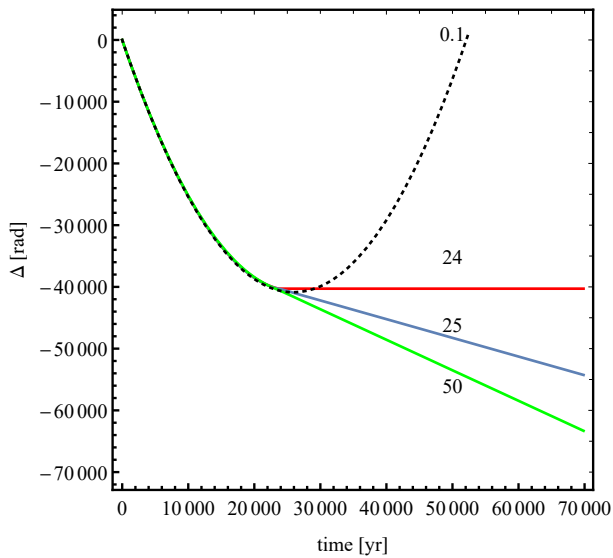}
 \caption{The time evolution of the phase angle $\Delta$  for systems with various coupling parameters $D$ (shown in the figure). At $t\lsim 20000$\,yr, all of the four curves are nearly degenerated.  For $D=0.1$, the coupling between the two inner binaries is weak, and the system just passes through the synchronization point $\dot \Delta=0$, globally approximated by the parabolic equation (\ref{pb}).  The system with $D=24$ is  captured into a synchronization state $\Delta \sim 40286$.   For $D=25$ and 50, the systems result in  drifting solutions $\dot \Delta\sim \rm  const$.  }
\end{figure}

\subsection{CAPTURE INTO SYNCHRONIZATION }

First, as a comparison to the drifting solutions, we discuss the numerical run with $D=24$ resulted in synchronization capture. 
In Fig. 3,  we plot its potential $V(\Delta)$  and the total energy $E(t)$ (red curve using Eq. (\ref{et})) around the synchronization capture. The red curve makes the first contact with the orange one (i.e. ${\dot \Delta}=0$) around $\Delta =-40287.8$,  and $\dot \Delta $ changes its sign form negative to positive.

Before this first contact, we  have ${\dot \Delta}<0$ and  the  product  $\lkk\cdots\rkk \times {\dot \Delta}$ in Eq. (\ref{et2}) has a  negative mean, reducing the  total energy $E(t)$. The wavy profile of the red curve in the bottom panel reflects the $\Delta$-dependence of the  mass-transfer factor  $\lkk\cdots\rkk$ in Eq. (\ref{et2}), mainly caused by the Newtonian angular momentum exchange.  Following the arguments in Appendix A, we can see that this wavy component is approximately proportional to  $ \sin2\Delta$.

  In Fig. 3, after the first contact, the phase angle $\Delta$  starts to oscillates in the potential (see the upper panel).  Now, the  mass transfer factor $\lkk\cdots\rkk$ in Eq. (\ref{et2}) has an oscillating component in the anti-phase with $\dot  \Delta$, efficiently decreasing the total energy $E(t)$ down to the bottom of the potential ($\cos2\Delta\sim -1$). Along the way, the GW luminosity  decreases significantly due  to the phase cancellation, as reported in Seto (2018) (see also a related explanation in \S 4). Although we do not provide the corresponding numerical results here, the primary binary  extracts the angular momentum of the secondary, realizing a parasitic relation with ${\dot m}_{\rm p2}=0$ at $t\gsim 3.8\times 10^4$\,yr (see Fig. 4 in Seto 2018 for a similar situation).

 In the four runs shown in  Fig. 2, except for the late stage of $D=24$, we always  have ${\dot m}_{\rm p2}<0$ and ${\dot m}_{\rm s2}<0$  simultaneously.  Under these two  inequalities,  the step function in Eq. (\ref{dm}) plays no role, and can be omitted, when interpreting  our  numerical calculations.

\begin{figure}
 \includegraphics[width=.95\linewidth]{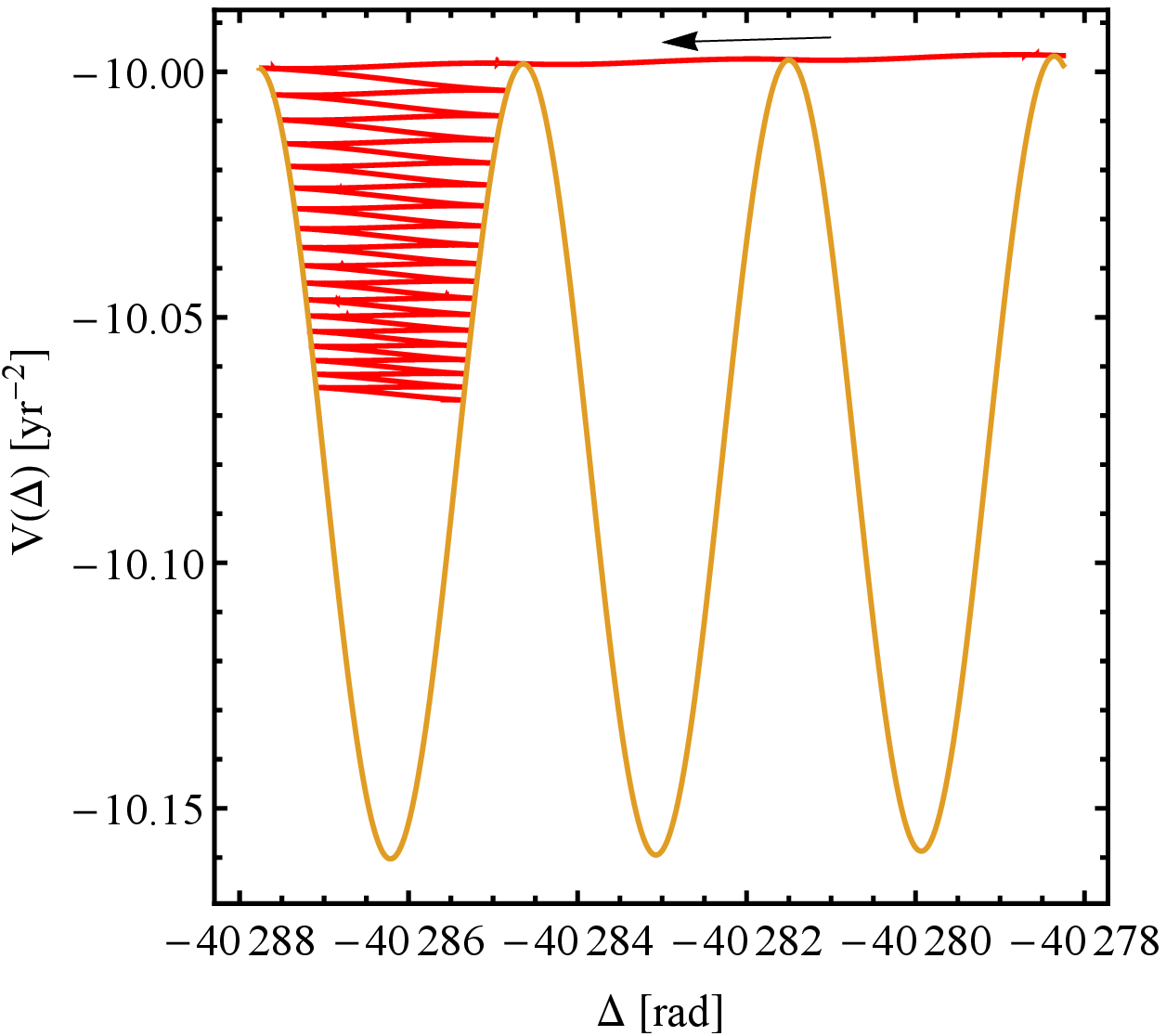}
 \includegraphics[width=.95\linewidth]{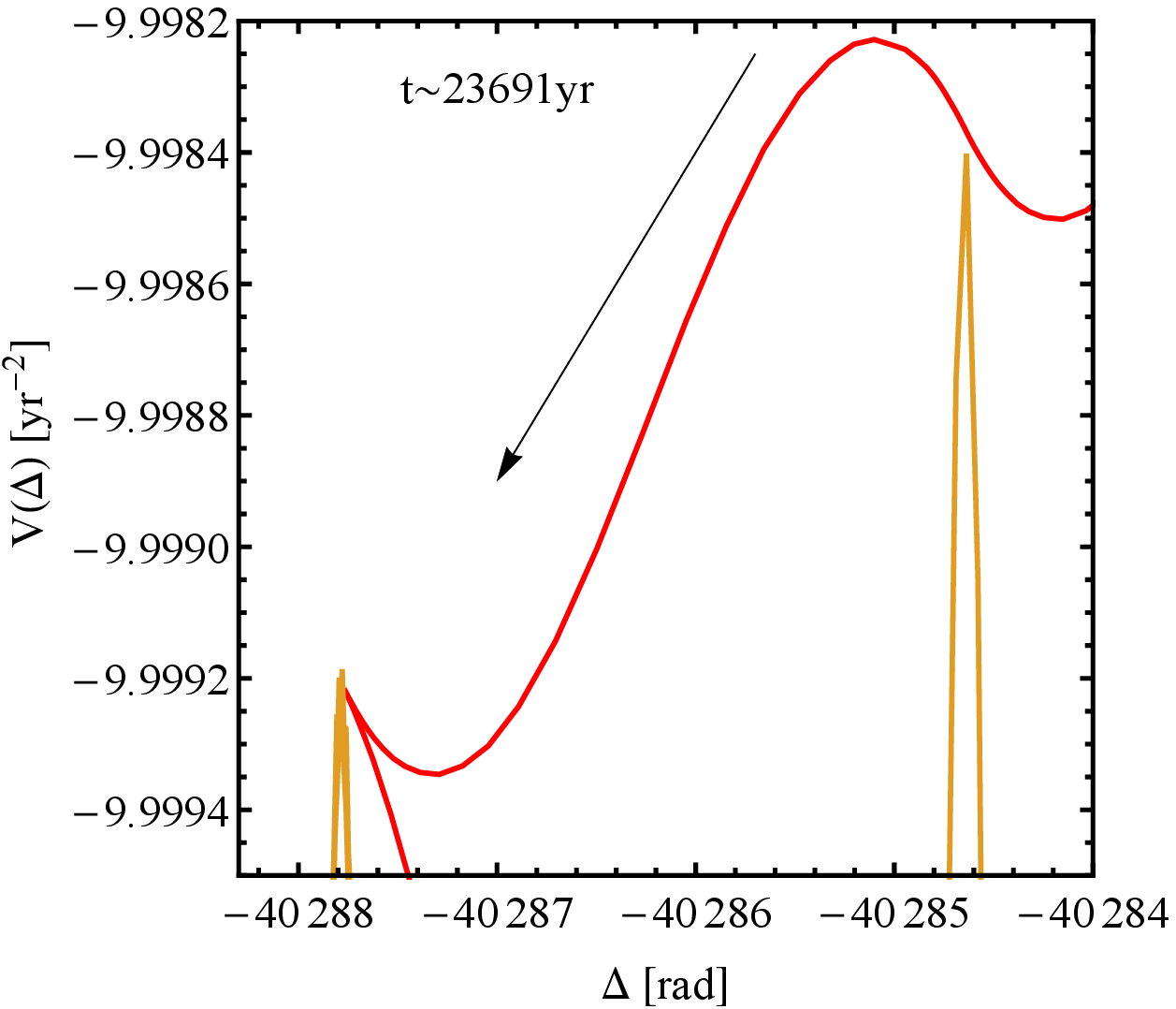}
 \caption{(Upper panel) The time evolution of the system with $D=24$ between  $t=23678$-24024\,yr around the synchronization capture. The red curve shows the total energy $E(t)={\dot \Delta}^2/2+V(\Delta)$, relative to the potential $V(\Delta)$ (orange curve).  This system is captured into the synchronization state $\Delta\simeq -40286.25$ (with $\cos2\Delta\simeq -1$).   (Lowe panel) An enlarged view of the upper panel around the first turning point at $t\sim 23691$\,yr. 
  }
\end{figure}

\subsection{DRIFTING SOLUTION}

Next, we discuss the run with $D=25$.  As shown in Fig. 2, at the late stage $t\gsim 30000$\,yr, this system shows a nearly constant drift  rate with the mean value 
\beq
A\equiv |{\bar {\dot \Delta}}|=0.303 \rm \,yr^{-1}.
\eeq

\begin{figure}
 \includegraphics[width=.88\linewidth]{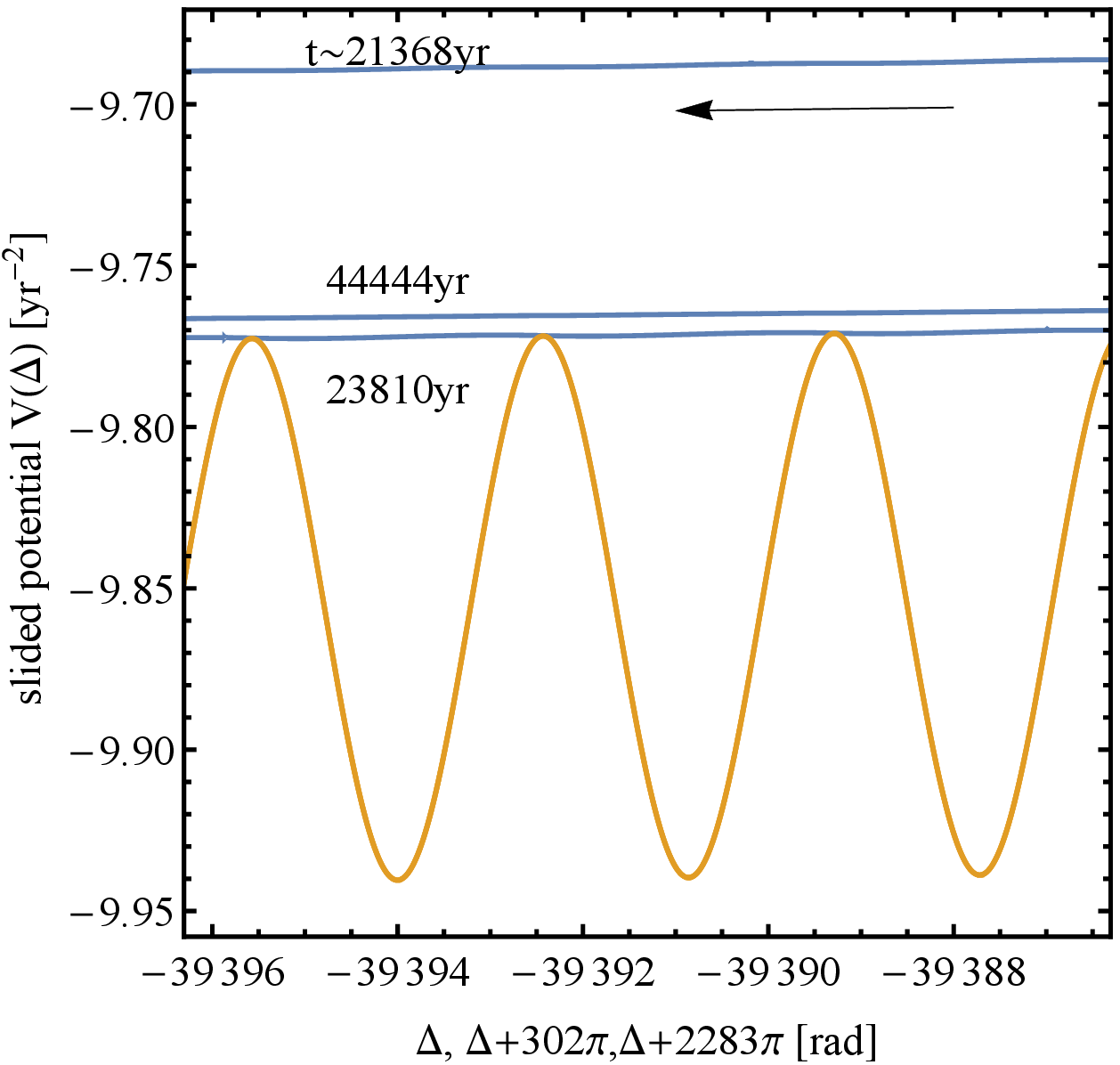}
 \includegraphics[width=.88\linewidth]{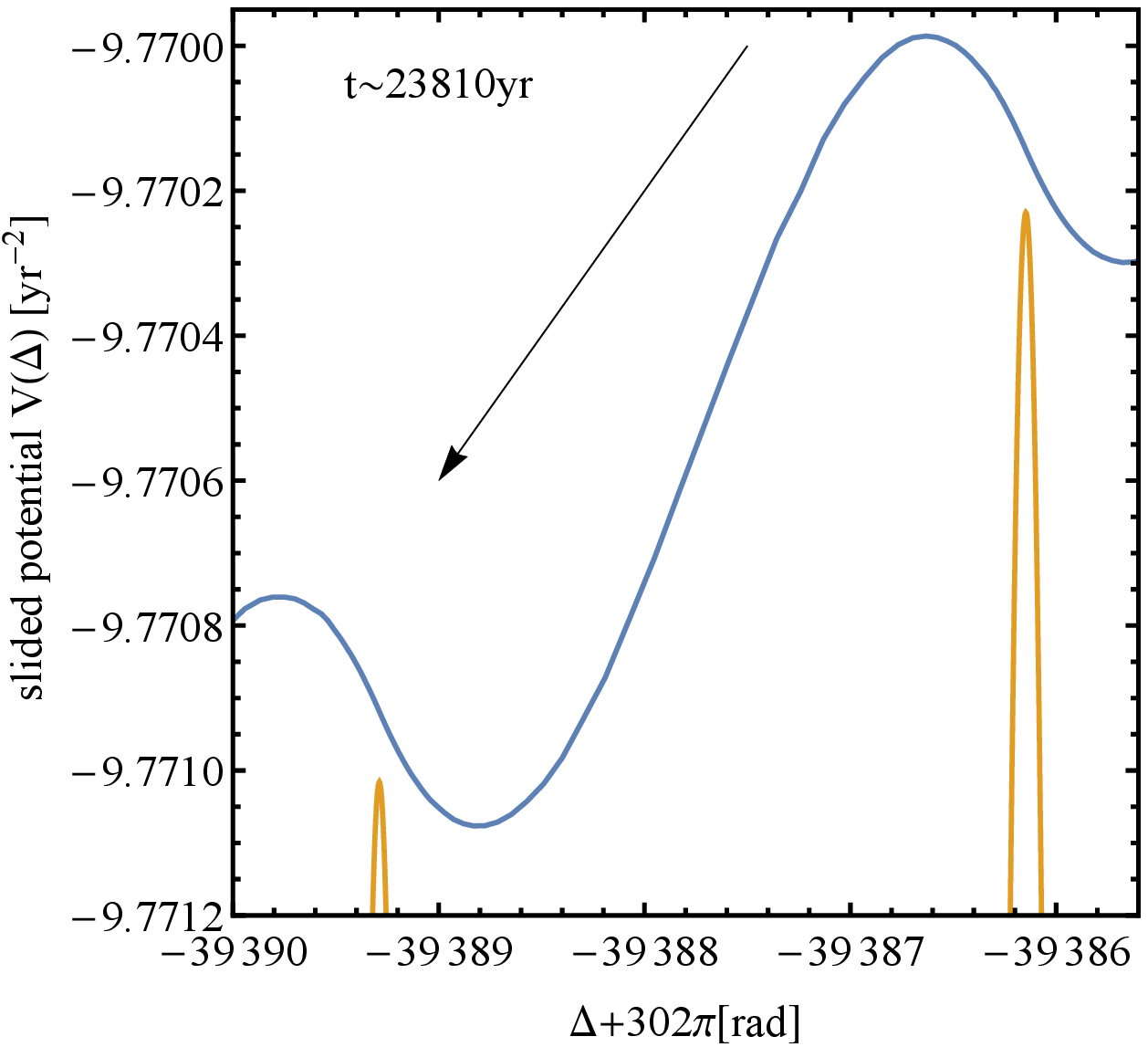}
 \caption{(Upper panel) The time evolution of the system with $D=25$ around three different epochs $t\sim21368, 23810$ and 44444\,yr. The potential $V(\Delta)$ (orange curve) is given for the total energy $E(t)$  shown  by the uppermost blue curve (labeled with 21368\,yr).  For the later two epochs, we appropriately slided both $E(t)$ and $V(\Delta)$, to compare with the first epoch.  (Lower panel) An enlarged view of Fig. 4 for the time evolution around 23810\,yr.  The blue curve does not contact with the orange curve, keeping ${\dot \Delta}<0$.}
\end{figure}

 In the upper panel of Fig. 4, we present the total energy $E(t)={\dot \Delta}^2/2+V(\Delta)$ (the upper blue line) and the potential $V(\Delta)$ (the orange curve) around $t\sim 21368$\,yr. At this relatively early stage, the system nearly keeps the initial acceleration ${\ddot \Delta}={\ddot \Delta}_{\rm iso}$, satisfying the parabolic equation (\ref{pb}) in the same way as other runs.
In  Fig. 4, the upper blue line has a slightly larger slope than that of the linear term $3n (F-1)\Delta /t_{\rm gw,s}$ of the potential $V(\Delta)$, gradually decreasing the kinetic energy $E(t)-V(\Delta)$.

In the upper panel of Fig. 4, we added $E(t)$ and $V(\Delta)$ around $t\sim 23810$\,yr where two curves experienced  the closest approach.  Since the potential  $V(\Delta)$ effectively has a repetitive shape  and only the relative position of the two curves are relevant for our study (showing $\dot \Delta^2/2$), we commonly slided  both $E(t)$ and $V(\Delta)$ in the horizontal and vertical directions, to directly compare with the situation at $t\sim 21368$\,yr mentioned earlier.   

Because of the poor resolution of the upper panel of Fig. 4 in the vertical direction, the blue line appears to contact with the orange curve.  But they are actually separated, as presented in the bottom panel. As in the case of the bottom panel of Fig. 3, we can see the wavy component $\psim \sin2\Delta$.

Finally, in the upper panel of Fig. 4,  during the drifting epoch around $t\sim 44444$\,yr, the energy $E(t)$ shows a clear offset from the potential (again appropriately slided in the horizontal and vertical directions).  This drifting solution can be regarded as a limit cycle sustained by the self-regulating mass transfer within two binaries (analytically examined in Appendix A).

In the upper panel of Fig. 4, comparing the chronological order of the three blue curves,  the total energy $E(t)$  does not change monotonically, relative to the potential $V(\Delta)$.   But it bounces back to the upper direction, before relaxing to the limit cycle.  This overshooting will become important in the next subsection. 

{For comparison, we also examine the transition point between the synchronization and the drifting solution, using different mass combinations  $(M_p,M_s)=(1.0M_\odot,0.95M_\odot)$ and $(1.0M_\odot,0.85M_\odot)$, still at $n=0.005\,{\rm s^{-1}}$.  The parameter $F\equiv (\mch_{\rm p}/\mch_{\rm s})^{5/3}$ becomes 1.037 and 1.121 respectively, and  we have transition at $D=5.9$ for the former and 59 for the latter.  If we increase the total mass difference between  two coupled  WD-WD binaries,  the intrinsic encounter speed  $\ddot{\Delta}_{\rm iso}\propto (F-1)$ increases, and we need a larger coupling parameter $D$  to keep the system in a drifting solution. Similarly,  for the coupling between a WD-WD binary and a WD-black hole binary, a much larger parameter $D$ would be required for the drifting solution, because of a larger encounter speed $\propto F-1$. }

\subsection{DIFFERENCE BETWEEN  $D=24$ AND $25$}
As shown in Fig. 2, we have  the distinct outcomes for $D=24$  and  25.  The former is captured into synchronization, but the latter  has a drifting solution. Here we briefly discuss  the structure of their boundary, mainly from  an interest in dynamical systems rather than from astronomical point of views.

One might imagine that, at $D\lsim 24$, we no longer have a corresponding drifting solution. However, considering the overshooting of the blue curves in Fig. 4,  it seems reasonable to presume that, even at $D\sim 24$, we still have a similar drifting solution, but the system in Fig. 3 was captured into synchronization state, by touching the potential curve and changing the sign of $\dot \Delta$ (as a result of the overshooting observed in Fig. 4).

 In order to clarify the existence of the drifting solution at $D\lsim 24$, we performed a numerical experiment, artificially changing the coupling parameter $D$ as a function of time 
\beq
D(t)=D_{\rm i}+(D_{\rm f}-D_{\rm i}) \lkk \exp\lmk \frac{t_{\rm t}-t}{t_{\rm d}} \rmk +1   \rkk^{-1} 
\eeq
with $D_{\rm i}=55$, $D_{\rm f}=15$, $t_{\rm t}=9.5\times10^4$\,yr and   $t_{\rm d}=3.2\times10^4$\,yr.  This function smoothly connects two values from $D\sim 55$ at $t\ll t_{\rm t}$ and $D\sim 15$ at $t\gg t_{\rm t}$.  Our intention behind this numerical experiment can be explained as follows.   

(i) We  initially relax the system to a drifting solution with the large coupling parameter $D=55$. 

(ii) Then, using a relatively long transition time-scale $t_d$ and suppressing the overshooting, we adiabatically lead the system down to $D=15$.

With this function $D(t)$, we could indeed realize a drifting solution even for $D=15$.  Therefore, the difference between Figs. 3 and 4 and is not caused by disappearance of a valid limit cycle, but by the effect of the  transient overshooting.

\section{Variable gravitational wave amplitude}
In this section, we discuss the time variation of gravitational wave amplitude, induced by the phase  drift.

Considering the aligned orientation of the two coupled binaries,  their quadrupole gravitational waves  are written as
\beqa
h_{\rm p}(t)&=&h F\cos[2\phi_{\rm p}(t)]=hF\cos[2\phi_{\rm s}(t)-2\Delta(t)],\label{hp1}\\
h_{\rm s}(t)&=&h \cos[2\phi_{\rm s}(t)].\label{hs1}
\eeqa
Here the amplitudes $h_{\rm p}$ and $h_{\rm s}$  depend on various geometrical parameters, but their explicit form is not important for the present arguments.  We also neglected the small Doppler effects induced by the outer orbital velocity. 
If the inter-binary separation $d$ is smaller than the gravitational wavelength $ \lambda$ as assumed in this paper, the total signal is effectively given by
\beq
h_{\rm total}(t)=h_{\rm p}(t)+h_{\rm s}(t).
\eeq

For the drifting solution,  the phase difference $\Delta(t)$ in Eq. (\ref{hp1}) changes much  more slowly than the inner orbital  angles $\phi_{\rm s}$. Therefore, due to the beat effect,  the total gravitational waveform changes its  amplitude as follows
\beq
{\cal A}_{\rm total}=h\lkk 1+F^2+2F\cos2\Delta(t)\rkk^{1/2}.\label{ampa}
\eeq
The amplitude ${\cal A}_{\rm total}$ has a positive interference ${\cal A}_{\rm total}\sim h(F+1)$ at  $\cos2\Delta(t)=1$, but has a cancellation  ${\cal A}_{\rm total} \sim h(F-1)$ at   $\cos2\Delta(t)=-1$.

In  this manner, the drifting solution would be interesting also from the viewpoint of gravitational wave  observation.  At the same time, we should notice that this amplitude variation is not merely an observational effect, but the intrinsic energy emission rate actually changes as $\propto {\cal A}_{\rm total}^2$. 

\begin{figure}
 \includegraphics[width=.95\linewidth]{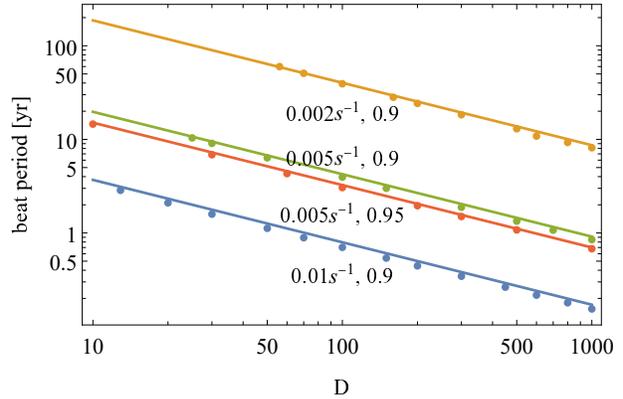}
 \caption{The beat periods $T_{\rm b}$ for the parameter  combinations  $(n\,[{\rm s^{-1}}],M_{\rm s}\,[M_\odot])=(0.002,0.90)$,  (0.005,0.90), (0.01,0.90) and (0.005,0.95) all with the fixed value  $M_p=1.0M_\odot$. The points are obtained from numerical experiments and the  solid lines are the analytical expressions given by Eq.  (\ref{ana}).  }
\end{figure}

If we simply put $\dot \Delta=$const,  ignoring its time modulation (see Eq. (A8)), the time averaged amplitude is  estimated to be  
\beq 
\lla {\cal A}^2_{\rm total}\rra^{1/2}=h(1+F^2)^{1/2} .\label{avr}
\eeq  We will use this expression later in \S 6.

\section{Beat period}
As discussed in the previous section, the drifting solution generates amplitude variation of gravitational waves due to a beat effect. 
 The beat period $T_{\rm b}$ is given by the mean drift rate $A$ as 
\beq
T_{\rm b}=\pi/A.
\eeq  For example, we have $T_{\rm b}\sim 10$\,yr for the system    discussed in \S 3.3 with $D=25$.  This period would be  suitable for observation by LISA (Amaro-Seoane et al. 2012).       
Considering these aspects, in this section, we specifically study the drift rate $A$ or equivalently the beat period $T_{\rm b}$.  Since our 
 model is a highly simplified one, we do  not necessarily take the actual numerical values too seriously. Instead, together with Appendix A, our discussion would help us to analytically understand the underlying  structure of the drifting solution.

So far, we have mainly examined the system  with $n\simeq0.005\,\rm s^{-1}$ and  $(M_{\rm p},M_{\rm s})=(1.0M_\odot,0.90M_\odot)$ for which drifting solutions are realized  at $D\gsim24$ as shown in Fig. 2. 
For comparison, we additionally study the following three cases;   $(n\,[{\rm s^{-1}}],M_{\rm s}\,[M_\odot])=(0.002,0.90)$, (0.01,0.90) and (0.005,0.95) all with $M_{\rm p}=1.0M_\odot$.  The minimum coupling parameters $D$ for the drifting solutions are 13, 56 and 5.9 respectively.

In Fig. 5, we show the  beat periods  $T_{\rm b}$ that were numerically obtained for various sets of the coupling parameters $D$ above the thresholds for the synchronization capture.  In this log-log plot, we can clearly observe the power-law relations $T_{\rm b}\psim D^{-2/3}$. 
Indeed, in Appendix A, we derive  the following analytical expression for the drift rate
\beq
A'=2^{-1/9} 3^{8/9} D^{2/3} (F-1)^{-1/3} n^{4/9}t_{\rm gw,s}^{-5/9}\label{ana}
\eeq
with the prime $\rq{}$ temporarily added to show the analytical counterpart to the original quantity $A$.  
For this derivation, we extracted the oscillating  components of the mass transfer rates and the semi-major axes, induced by the Newtonian torque.  Then we solve the drift rate $A\rq{}$ by using the energy balance equation (\ref{et2}). 
In Fig. 5, we added the analytical estimation $T_{\rm b}=\pi/A\rq{}$ as four curves. It well reproduces the numerical results.

Next, we briefly  discuss the long-term evolution of a drifting system with a given  inter-binary separation $d$.  The evolution timescale is approximately given by $t_{\rm gw,p}$ in Eq. (\ref{tgw}). Using the relation (\ref{dom}) for the donor mass, we have $t_{\rm gw,p}\psim n^{-11/3}$. Then, from Eqs. (\ref{defd}) and (\ref{ana}), we have $A\rq{}\psim d^{-10/3} n^{-23/27}$. 
For the specific  mass parameters $(M_{\rm p},M_{\rm s})=(1.0M_\odot,0.90M_\odot)$, we obtain the beat period 
\beq
T_{\rm b}=10 (d/{\rm 0.13AU})^{10/3} (n/0.005{\rm s^{-1}})^{23/27}{\rm yr}.
\eeq 

\section{observation with LISA}

\begin{figure}
 \includegraphics[width=.95\linewidth]{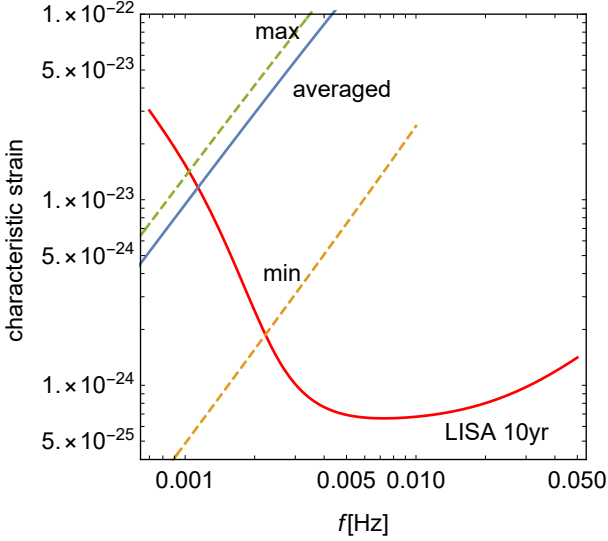}
 \caption{The three  characteristic strain amplitudes of a drifting system with  the total masses $(M_{\rm p},M_{\rm s})=(1.0M_\odot,0.90M_\odot)$ and the distance  $D_L=5$kpc.  The upper dashed line is the maximum amplitude at $\cos2\Delta =1$, and the bottom dashed line  is the minimum one at $\cos2\Delta =-1$. The blue solid line is the time averaged amplitude.  The effective noise levels $\sqrt{S_n(f) T^{-1}_{\rm obs}}$ of LISA  is given for the observation period $T_{\rm obs}=10$yr, and contains the Galactic confusion noise. }
\end{figure}

In this section, we discuss observation of a drifting four-body system with LISA.   Here, instead of the orbital angular velocity $n$, we use the gravitational wave frequency $f=n/\pi$. 

For an isolated circular binary, after taking its direction and orientation averages, the effective strain amplitude $h$ is given by
\beq
h=\frac{8 (G\mch)^{5/3} \pi^{2/3}f^{2/3}}{5^{1/2}c^4D_L}
\eeq
with the chirp mass $\mch$ and the binary distance $D_L$ (see e.g. Robson, Cornish \& Liu 2019).

As shown in Eq. (\ref{ampa}),  a  drifting system changes its gravitational wave amplitude ${\cal A}_{\rm total}$ between  $h(F-1)$ and  $h(F+1)$ with the time averaged value $h(F^2+1)^{1/2}$ given  in Eq. (\ref{avr}).    

In Fig. 6, we plot the three  amplitudes for the model parameters $(M_{\rm p},M_{\rm s})=(1.0M_\odot,0.90M_\odot)$ and  $D_L=5$kpc.
In this plot, the factor $F\equiv (\mch_{\rm p}/\mch_{\rm s})^{5/3}$  depends very weakly on  $f$. In fact, when the donor masses are much smaller than  accreter masses, we have $F\simeq (M_{\rm p}/M_{\rm s})^{2/3}\sim$const. For the present model parameters, we get $F\sim 1.05$, and  the amplitude changes by a factor of $(F+1)/(F-1)\sim 40$ during the single beat period.

If the beat period $T_{\rm b}$ is  smaller than the observation period $T_{\rm obs}$, the optimal  signal-to-nose ratio of the emitted waves can be evaluated with the averaged amplitude as
\beq
SNR\simeq \frac{h(F^2+1)^{1/2}}{\sqrt{S_n(f)T^{-1}_{\rm obs}}}.\label{snr}
\eeq
Here $S_n(f)$ is the standard strain noise spectrum of LISA and  defined in units of $[\rm Hz^{-1}]$ (Robson, Cornish \& Liu 2019).  In Fig. 6, for $T_{\rm obs}=10$\,yr,   we show the effective noise levels $\sqrt{S_n(f) T^{-1}_{\rm obs}}$. Applying Eq. (\ref{snr}) to our model parameters above,   we have $SNR=67$ at $f=3.2$mHz and 12 at 2mHz.


{
 For detecting the amplitude modulated waves  with LISA, we consider the following  two step data analysis. The first step is selecting candidates of drifting systems,  using relatively short-term data.   The next step is the follow-on examination of the candidates whether they have  long-term amplitude modulations.  For the first step, we can perform  a  matched filtering analysis,  approximately using the short-term templates  made for standard nearly monochromatic isolated binaries.   For example, in the 3.2m\,Hz case above, we can get $SNR\sim8$ typically in the period $\rm 10\, yr \times (8/67)^2=0.14$\,yr that could be much smaller than the beat period. Another method for the first-step candidate selection is  a search for localized power in a narrow frequency interval  (see Cornish \& Larson 2003 for the Doppler demodulation).  In reality, the candidates after the first step will be dominated by  simple isolated binaries.  But, after the second step,  we might identify a small number of drifting systems. 
}

\section{summary and Discussion}
In this paper, using a very simple model  based on  Paczy{\'n}ski (1967) and Paczy{\'n}ski \& {Sienkiewicz} (1972), we examine evolution of coupled dual mass-transferring WD binaries around the synchronization point $n_{\rm p}\sim n_{\rm s}$. We find that, in a strongly coupled configuration (i.e. short mutual distance), the system can asymptotically settle into a drifting solution as a limit cycle.  This state is remarkably different from a synchronization capture realized in less strong coupling (Seto 2018).

{
Considering \lq\lq{}stability\rq\rq{} of the drifting solution against small perturbation as shown in Fig.  4, we can qualitatively expect that such solutions would be maintained to some extent, even adding small corrections to our simple model.  But,  to better understand what actually happens around the synchronization point, we need to quantitatively examine various  physical effects that are not included in the present model. For example, our formulation  is based on the balances of angular momenta.  But, 
for each binary, as shown in Eq. (3), we only considered  the orbital angular momentum essentially for two point masses.  In reality, the angular momentum is partially stored in the spin rotations of the accreter and possibly in its accretion  disk, or might be lost from the four-body system due to a mass loss (see e.g. Marsh et al. 2004; Gokhale et al. 2007; Solheim 2018). These corrections also affect the response of orbital angular velocity to externally added torque, and could play interesting roles  for the dynamical couplings.  Meanwhile, we just included the resonant torques for the inter-binary interaction. But other short-term torques might disturb the ordered structures studied in this paper (see e.g. Murray \& Dermott 1999 for the effects of non-resonant terms).  
In any case, our study here  is far from complete, and additional effects are worth considering. }


The drifting solution generates amplitude variation of emitted gravitational waves, due to a beat effect. Depending on model parameters,   the beat period could be 1-10\,yr and a large amplitude variation might be actually observed by LISA.  In this respect, we might detect other associated signatures encoded in gravitational waveform such as a small phase modulation caused by the outer orbital motion.

In this paper, we concentrate our study around the synchronization point $n_{\rm p}\sim n_{\rm s}$ where our formulation is applicable. It would be also interesting to discuss other evolutionary stages, especially possible pathways to forming  strongly coupled four-body systems as considered in this paper. We left these issues as future works.

\section*{Acknowledgements}
the author would like to thank the reviewer  for  valuable comments on the manuscript. 
 This work is supported by JSPS Kakenhi Grant-in-Aid for Scientific Research
 (Nos.~15K65075, 17H06358 and 19K03870).

\section*{DATA AVAILABILITY}
The data underlying this article will be shared on reasonable request to the corresponding author.

\bibliographystyle{mn2e}


\appendix

\section{Analytical evaluation for the drift rate}
In this appendix, we derive an analytical expression for the drift rate $A \equiv |{\bar {\dot \Delta}}|$.  This derivation would be useful also to understand the underlying structure of the drifting solution. We assume $D\gg 1$ and put $F=1$, except for the combination $(F-1)$. 

During the drift,  the two binaries exchange angular momenta with the Newtonian torque $T_{\rm p}=-T_{\rm s}\propto \sin2\Delta$ (see Eq. (\ref{new})) at the mean angular speed $2A$. Our basic strategy here is (i) to derive relations for the oscillating components of  $(\dot{m}_{\rm p2},\dot{m}_{\rm s2})$ and $(a_{\rm p},a_{\rm s})$ induced by the exchange, and (ii) to subsequently estimate the drift rate $A$ by using the averaged energy variation rate.

 To begin with, for the mass transfer rate $\dot{m}_{\rm p2}$, we decompose the nearly constant (DC) part  ${\bar {\dot m}}_{\rm p2}$  and the oscillating (AC) part $\delta {\dot m}_{\rm p2}$ as follows
\beq
{ {\dot m}}_{\rm p2}={\bar {\dot m}}_{\rm p2}+\delta {\dot m}_{\rm p2}.
\eeq
Similarly, we separate the inner semi-major axis into the smooth part ${\bar{a}}_{\rm p}$  and the oscillating part $\delta a_{\rm p}$ 
\beq
a_{\rm p}={\bar{a}}_{\rm p}+\delta a_{\rm p}.
\eeq

Next, by perturbatively expanding Eq. (\ref{dm}) and using Eq. (\ref{iso}), we obtain the following expression 
\beq
\frac{\delta { {\dot m}}_{\rm p2}}{m_{\rm p2}}=6n  \frac{\delta {a}_{\rm p}}{a_{\rm p}} \lmk \frac{-{\bar{\dot m}}_{\rm p2}}{2nm_{\rm p2}} \rmk^{2/3}=6n  \frac{\delta {a}_{\rm p}}{a_{\rm p}} \lmk \frac{3}{4t_{\rm gw,s} n} \rmk^{2/3}.
\eeq
Here we neglected the contribution of $\delta m_{\rm p2}$ on the right-hand side of Eq. (\ref{iso}) (as justified shortly)
 and also dropped the step function (as already commented after Eq. (\ref{dm})). For the magnitudes of oscillation amplitude , we have
\beq
\frac{\delta { {\dot m}}_{\rm p2}}{m_{\rm p2}}\sim  \frac{\delta {\dot a}_{\rm p}}{a_{\rm p}} \lkk  \frac{3n}{A}\lmk \frac{3}{4t_{\rm gw,s} n} \rmk^{2/3}\rkk \ll \frac{\delta {\dot a}_{\rm p}}{a_{\rm p}}  .
\eeq
In the above relation, we use the fact that the factor $\lkk\cdots\rkk$ is much smaller than unity  for  the actual numerical data $A$ shown in Fig.  5.
After taking time integration, we can expect a similar hierarchy for $\delta m_{\rm p2}/m_{\rm p2}\ll \delta a_{\rm p}/a_{\rm p}$, justifying the expansion of  Eq. (A3) only with $\delta a_{\rm p}/a_{\rm p}$. From Eqs. (\ref{djp2}) and (A4), we have 
\beq
 \frac{\delta {\dot a}_{\rm p}}{a_{\rm p}} =\frac{2D}{t_{\rm gw,s}} \sin(2\Delta).
\eeq
For the secondary binary, we can show similar relations
\beq
\frac{\delta { {\dot m}}_{\rm s2}}{m_{\rm s2}}=6n  \frac{\delta {a}_{\rm s}}{a_{\rm s}} \lmk \frac{3}{4t_{\rm gw,s} n} \rmk^{2/3},
\eeq
\beq
 \frac{\delta {\dot a}_{\rm s}}{a_{\rm s}} =-\frac{2D}{t_{\rm gw,s}} \sin(2\Delta).
\eeq
Integrating Eq. (\ref{b1}) after using Eqs. (A5) and (A7),  we have
\beq
\dot{\Delta}=-A-\frac{3nD}{t_{\rm gw,s}A}\cos(2\Delta).
\eeq
Here we applied the condition $A \equiv |{\bar {\dot \Delta}}|$ for determining the integral constant, and put $\int^t \sin2\Delta dt=-(2A)^{-1}\cos2\Delta$ for the correction term.
From Eqs. (A3) and (A6), keeping the term relevant for the arguments below,  we have
\beq
\frac{\delta {\dot m}_{\rm p2}}{m_{\rm p2}}-\frac{\delta {\dot m}_{\rm s2}}{m_{\rm s2}}=C\cos(2\Delta)
\eeq
with
\beq
C=12 \frac{nD}{t_{\rm gw,s}A} \lmk \frac{3}{4t_{\rm gw,s} n} \rmk^{2/3}.
\eeq

Now, we solve the unknown parameter $A$, by evaluating the long-term energy variation rate $\dot E$ in two different  ways. First, using the steady drift of potential energy (through its linear term), we have 
\beq
\overline{\dot E}= {\overline{\frac{\p V}{\p \Delta}}\bar{\dot \Delta}} =  -\frac{3An(F-1)}{t_{\rm gw,s}}
\eeq
Meanwhile, using Eq. (\ref{et2}) and taking the time average of the following combination 
\beq
3n \lmk \frac{\delta {\dot m}_{\rm p2}}{m_{\rm p2}}-\frac{\delta {\dot m}_{\rm s2}}{m_{\rm s2}}\rmk {\dot \Delta},
\eeq
we have
\beq
\bar{\dot E}=-\frac{9Dn^2C}{2At_{\rm gw,s}}.
\eeq
Here, in the asymptotic stage, the constant parts ($\bar{\dot m}_{\rm p2}, \bar{\dot m}_{\rm s2})$ are expected to be almost canceled in Eq. (\ref{et2}) and we only kept the oscillating (anti-phase) parts, ignoring the small parameter $q_{\rm p}\sim q_{\rm s}\ll 1$. Matching Eqs. (A11) and (A13), we finally obtain
\beq
A=2^{-1/9} 3^{8/9} D^{2/3} (F-1)^{-1/3} n^{4/9}t_{\rm gw,s}^{-5/9}.
\eeq

\end{document}